\documentclass[12pt]{article}
\usepackage{graphicx}
\usepackage{hyperref}
\usepackage{orcidlink}

\title{\textit{Hyper-py}: HYbrid Photometry and Extraction Routine in PYthon}

\author{
Alessio Traficante$^{1}$\orcidlink{0000-0003-1665-6402}\thanks{Corresponding author: \texttt{alessio.traficante@inaf.it}}, Fabrizio De Angelis$^{1}$, Alice Nucara$^{1}$, \\
Milena Benedettini$^{1}$ \\
\\
$^{1}$INAF-IAPS, Via Fosso del Cavaliere, 100, 00133 Rome, Italy
}

\date{}

\begin{document}

\maketitle

\section{Summary}

Source extraction and photometry of compact objects are key tasks in observational astronomy. Numerous tools have been developed to tackle the complexities of astronomical data, especially for precise background estimation and source deblending, which are essential for reliable flux measurements across wavelengths (e.g., Cutex: \cite{Molinari11}; getsources: \cite{Menshinkov12}; Fellwalker: \cite{Berry15}; Astrodendro). These challenges are particularly significant in star-forming regions, best observed in the far-infrared (FIR), sub-millimeter, and millimeter bands, where cold, dense compact sources emit strongly. To address this, several software packages have been designed for handling the structured backgrounds and blended sources typical of observations by instruments like Herschel (70–500 $\mu$m) and ALMA (1–3 mm). These tools differ in detection and flux estimation approaches. Within this context, we developed HYPER (HYbrid Photometry and Extraction Routine, \cite{Traficante15}), originally implemented in IDL, aiming to deliver robust and reproducible photometry of compact sources in FIR/sub-mm/mm maps. HYPER combines source detection via high-pass filtering, background estimation through local polynomial fitting, and source modeling with 2D elliptical Gaussians, simultaneously fitting multiple Gaussians to deblend overlapping sources.

Aperture photometry in HYPER is performed on background and companion subtracted images, using footprints defined by the source’s 2D Gaussian models, ensuring robust flux measurements in crowded or structured environments \cite{Traficante15,Traficante23}. The hybrid approach combines parametric Gaussian modeling with classical aperture photometry. Here, we present \textbf{\textit{Hyper-Py}}, a fully restructured and extended Python implementation of HYPER. \textit{Hyper-Py} preserves the original logic while offering improvements in performance, configurability, and background modeling capabilities, making it a flexible modern tool for source extraction and photometry across diverse datasets. Notably, \textit{Hyper-Py} enables background estimation and subtraction across individual slices of 3D datacubes, allowing consistent background modeling along the spectral axis for line or continuum studies in spectrally resolved observations.

\section{Statement of need}

\textit{Hyper-Py} is an open-source Python package freely available to the community. This new implementation builds upon and improves the original IDL HYPER by incorporating several major advancements:

\subsection*{Parallel execution for multi-map analysis}
\textit{Hyper-Py} employs built-in parallelization where each input map is independently assigned to a processing core on multi-core systems. This allows concurrent execution of the complete photometric pipeline on different maps simultaneously. This parallel framework dramatically increases computational efficiency without altering individual map results.

\subsection*{Native support for FITS datacubes}
The software treats each slice along the third axis as an independent 2D map, compatible with parallel processing, allowing simultaneous background subtraction per slice. The output is a 3D background cube matching the input cube’s shape, configurable for targeted regions or full spatial coverage through a user-friendly configuration file. This capability provides flexibility for both line-specific and broader continuum background modeling.

\subsection*{Improved source detection reliability}
Source detection has been enhanced with a robust sigma-clipping algorithm that iteratively estimates the root mean square (\textit{rms}) noise of input maps, excluding outliers to characterize background fluctuations accurately—even with bright sources or structures present. This \textit{rms} serves as a threshold reference for detecting compact sources exceeding a configurable significance level ($n_\sigma \times \textit{rms}$), settable via the config file. Such refinement increases detection reliability and reproducibility across heterogeneous datasets.

\subsection*{Advanced background estimation strategy}
Unlike the original IDL implementation \cite{Traficante15}, which modeled background separately from source fitting, \textit{Hyper-Py} supports multiple statistical fitting techniques, least-squares, Huber, and Theil–Sen regressions, applied to masked cutouts around each source. Least-squares performs well in regions dominated by Gaussian noise; Huber regression balances L2 and L1 losses to reduce outlier effects via a tunable parameter $\epsilon$ (\texttt{huber\_epsilons} in the config file); and Theil–Sen is a non-parametric, robust method ideal for non-Gaussian noise or contamination. When multiple methods are enabled, \textit{Hyper-Py} selects the best model by minimizing residuals, ensuring accurate background reconstruction even with gradients or faint extended emission. Furthermore, an optional joint fit of background and 2D Gaussian models with L2 (ridge) regularization stabilizes fits in regions with strong gradients, preventing background overfitting at the expense of source flux.

\subsection*{Gaussian plus background model optimization strategy}
\textit{Hyper-Py} utilizes the Levenberg–Marquardt algorithm through the \texttt{lmfit} package’s \texttt{least\_squares} minimizer, allowing control over the cost function’s residual weighting by selecting different loss models. The default “cauchy” loss diminishes outlier influence, improving robustness to data artifacts, unmasked sources, or non-Gaussian noise. Alternatives like “soft\_l1” and “huber” are also available for specific dataset optimization.

\subsection*{Model selection criteria for background fitting}
Background model selection criteria are configurable, offering Normalized Mean Squared Error (NMSE), reduced chi-squared ($\chi^2_\nu$), or Bayesian Information Criterion (BIC) via the config file. NMSE is default due to its robust, scale- and weighting-independent nature, ideal under varying masking or pixel weighting. Reduced chi-squared depends on noise models and pixel counts, potentially biasing selection. BIC penalizes model complexity, favoring simplicity when residuals are comparable. These options allow users to select criteria best suited to their scientific aims and data noise properties.

\subsection*{Improved user configurability}
\textit{Hyper-Py} is designed to be more user-friendly, featuring a clear and well-documented configuration file. This allows users to adapt the full photometric workflow to a wide range of observational conditions and scientific goals by modifying only a minimal set of parameters.

\subsection*{Performance assessment}
We assessed \textit{Hyper-Py} performance using extensive simulations. Starting from a noise-only map based on real ALMA data, we generated two maps with reference headers and superimposed varying backgrounds plus 500 synthetic 2D Gaussian sources. These sources mimic real compact objects with integrated fluxes spanning 8–20 times the map rms (peak fluxes $\sim$1–1.5 $\times$ \textit{rms}) and FWHM sizes of 0.5–1.5 times the beam size to include both unresolved and moderately extended sources \cite{Elia21}. Random position angles and a minimum 30\% overlap ensured realistic blending, thus providing a rigorous test of the code under realistic and challenging conditions. We compared the original IDL HYPER and \textit{Hyper-Py} under equivalent configurations, with the latter benefiting from improved background estimation, optional regularization, and parallel processing. The key results are presented in Table 1, detailing differences in source identification and false positives between the codes.

\begin{table}[h]
\centering
\begin{tabular}{r l r r r r}
\hline
Catalog & Source Type & Total & Matched & False & False Percentage \\
\hline
1 & \textit{Hyper-Py} & 500 & 490 & 4 & 0.8\% \\
1 & HYPER (IDL) & 500 & 493 & 73 & 12.9\% \\
2 & \textit{Hyper-Py} & 500 & 487 & 4 & 0.8\% \\
2 & HYPER (IDL) & 500 & 487 & 46 & 8.6\% \\
\hline
\end{tabular}
\caption{Comparison between Hyper-Py and HYPER (IDL) performance on simulated maps.}
\end{table}

In addition, Figure 1 and Figure 2 show the differences between the peak fluxes and the integrated fluxes of the sources with respect to the reference values of the simulated input sources as estimated by \textit{Hyper-Py} and HYPER, respectively.

\textit{Hyper-Py} is freely available for download via its \href{https://github.com/Alessio-Traficante/hyper-py}{GitHub repository}\footnote{https://github.com/Alessio-Traficante/hyper-py}, and can also be installed using \texttt{pip}, as described in the accompanying README file available in the repository.

\begin{figure}[h]
\centering
\includegraphics[width=0.8\linewidth]{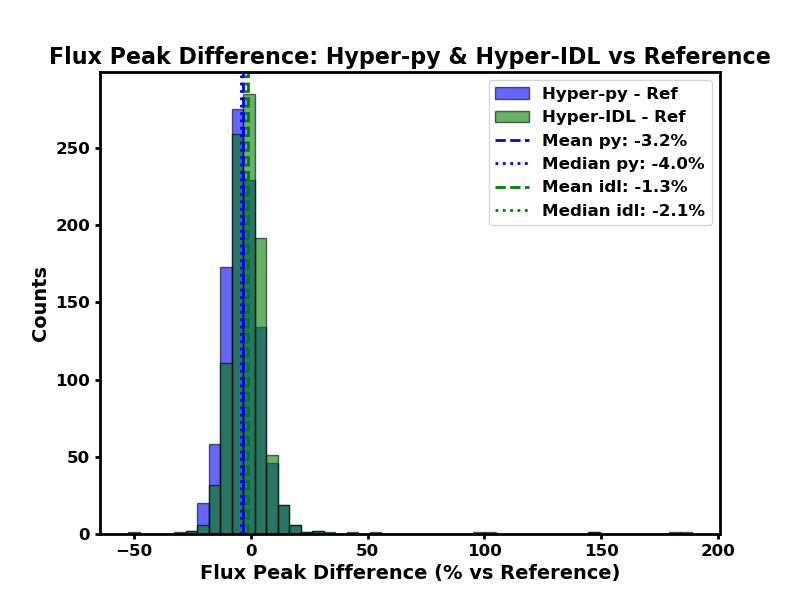}
\caption{Histogram of the differences between the peak fluxes of the sources as recovered by \textit{Hyper-Py} and HYPER, respectively, with respect to the reference values of the simulated input sources.}
\end{figure}

\begin{figure}[h]
\centering
\includegraphics[width=0.8\linewidth]{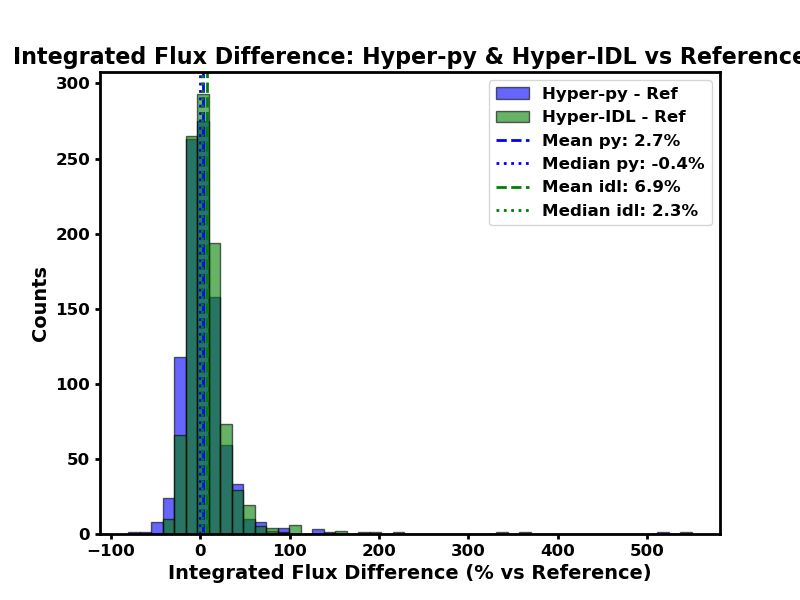}
\caption{Histogram of the differences between the integrated fluxes of the sources as recovered by \textit{Hyper-Py} and HYPER, respectively, with respect to the reference values of the simulated input sources.}
\end{figure}

\bibliographystyle{unsrt}
\bibliography{paper}

\begin{thebibliography}{1}

\bibitem{Molinari11}
S.~{Molinari}, E.~{Schisano}, F.~{Faustini}, M.~{Pestalozzi}, A.~M. {di Giorgio}, and S.~{Liu}.
\newblock {Source extraction and photometry for the far-infrared and sub-millimeter continuum in the presence of complex backgrounds}.
\newblock {\em A$\&$A}, 530:A133, June 2011.

\bibitem{Menshinkov12}
A.~{Men'shchikov}, Ph. {Andr{\'e}}, P.~{Didelon}, F.~{Motte}, M.~{Hennemann}, and N.~{Schneider}.
\newblock {A multi-scale, multi-wavelength source extraction method: getsources}.
\newblock {\em A$\&$A}, 542:A81, June 2012.

\bibitem{Berry15}
D.~S. {Berry}.
\newblock {FellWalker-A clump identification algorithm}.
\newblock {\em Astronomy and Computing}, 10:22--31, April 2015.

\bibitem{Traficante15}
A.~Traficante, G.~A. Fuller, N.~Peretto, and J.~E. Pineda.
\newblock High‐resolution maps of star‐forming regions - i. the hyper algorithm for compact source extraction.
\newblock {\em MNRAS}, 451(4):4086--4103, 2015.

\bibitem{Traficante23}
A.~{Traficante}, B.~M. {Jones}, A.~{Avison}, G.~A. {Fuller}, M.~{Benedettini}, D.~{Elia}, S.~{Molinari}, N.~{Peretto}, S.~{Pezzuto}, T.~{Pillai}, K.~L.~J. {Rygl}, E.~{Schisano}, and R.~J. {Smith}.
\newblock {The SQUALO project (Star formation in QUiescent And Luminous Objects) I: clump-fed accretion mechanism in high-mass star-forming objects}.
\newblock {\em MNRAS}, 520(2):2306--2327, April 2023.

\bibitem{Elia21}
D.~{Elia}, M.~{Merello}, S.~{Molinari}, E.~{Schisano}, A.~{Zavagno}, D.~{Russeil}, P.~{M{\`e}ge}, P.~G. {Martin}, L.~{Olmi}, M.~{Pestalozzi}, R.~{Plume}, S.~E. {Ragan}, M.~{Benedettini}, D.~J. {Eden}, T.~J.~T. {Moore}, A.~{Noriega-Crespo}, R.~{Paladini}, P.~{Palmeirim}, S.~{Pezzuto}, G.~L. {Pilbratt}, K.~L.~J. {Rygl}, P.~{Schilke}, F.~{Strafella}, J.~C. {Tan}, A.~{Traficante}, A.~{Baldeschi}, J.~{Bally}, A.~M. {di Giorgio}, E.~{Fiorellino}, S.~J. {Liu}, L.~{Piazzo}, and D.~{Polychroni}.
\newblock {The Hi-GAL compact source catalogue - II. The 360{\textdegree} catalogue of clump physical properties}.
\newblock {\em MNRAS}, 504(2):2742--2766, June 2021.

\end{thebibliography}

\end{document}